\documentclass[10pt]{iopart}

\usepackage{graphicx}
\usepackage{dcolumn}
\usepackage{bm}
\usepackage{color}

\begin{document}

\title[Alfven Eigenmodes in TJ-II]{Analysis of Alfven Eigenmodes destabilization by energetic particles in TJ-II using a Landau-closure model}


\author{J. Varela}
\ead{rodriguezjv@ornl.gov}
\address{Oak Ridge National Laboratory, Oak Ridge, Tennessee 37831-8071}
\author{D. A. Spong}
\address{Oak Ridge National Laboratory, Oak Ridge, Tennessee 37831-8071}
\author{L. Garcia}
\address{Universidad Carlos III de Madrid, 28911 Leganes, Madrid, Spain}

\date{\today}

\begin{abstract}
Alfv\' en Eigenmodes (AE) can be destabilized by energetic particles in neutral beam injection (NBI) heated plasmas through inverse Landau damping and couplings with gap modes in the shear Alfv\' en continua. We describe the linear evolution of the poloidal flux and the toroidal component of the vorticity in a full 3D system using the reduced MHD equations, density and parallel velocity moments for the energetic particles as well as the geodesic acoustic wave dynamics. A closure relation adds the Landau damping and resonant destabilization effects in the model. We apply the model to study the Alfv\' en modes stability in TJ-II, performing a parametric analysis in a range of realistic values of energetic particle $\beta$ ($\beta_{f}$), ratios of thermal/Alfv\' en velocities ($V_{th}/V_{A0}$), energetic particle density profiles and toroidal modes ($n$) including toroidal and helical couplings. The study predicts a large helical coupling between different toroidal modes and the destabilization of helical Alfv\' en Eigenmodes (HAE) with frequencies similar to the AE activity measured in TJ-II, between $50 - 400$ kHz. The analysis has also revealed the destabilization of GAE (Global Alfv\' en Eigenmodes), TAE (Toroidal Alfv\' en Eigenmodes) and EPM (Energetic Particle Modes). For the modes considered here, optimized TJ-II operations require a $\rlap{-} \iota$ profile in the range of $[0.845 , 0.979]$ to stabilize AEs in the inner and middle plasma. AEs in the plasma periphery cannot be fully stabilized, although for a configuration with $\rlap{-} \iota = [0.945 , 1.079]$, only $n=7,11,15$ AE are unstable with a growth rate $4$ times smaller compared to the standard $\rlap{-} \iota = [1.54,1.68]$ case and a frequency of $100$ kHz. We reproduce the frequency sweeping evolution of the AE frequency observed in TJ-II as the $\rlap{-} \iota$ profile is varied. The AE frequency sweeping is caused by consecutive changes of the instability dominant modes between different helical families.
\end{abstract}

%
%
%
%
%

\pacs{52.35.Py, 52.55.Hc, 52.55.Tn, 52.65.Kj}

\vspace{2pc}
\noindent{\it Keywords}: Stellarators, MHD, AE, energetic particles

This manuscript has been authored by UT-Battelle, LLC under Contract No. DE-AC05- 00OR22725 with the U.S. Department of Energy. The United States Government retains and the publisher, by accepting the article for publication, acknowledges that the United States Government retains a non-exclusive, paid-up, irrevocable, world-wide license to publish or reproduce the published form of this manuscript, or allow others to do so, for United States Government purposes. The Department of Energy will provide public access to these results of federally sponsored research in accordance with the DOE Public Access Plan (http://energy.gov/downloads/doe-public-access-plan).

\maketitle

\ioptwocol

\section{Introduction \label{sec:introduction}}

The transport of fusion produced alpha particles, energetic hydrogen neutral beams and particles heated using ion cyclotron resonance heating (ICRF) is affected by the energetic particle driven instabilities \cite{1,2,3}, potentially leading to a drop of the operation performance in devices as TFTR, JET and DIII-D tokamaks or LHD and W7-AS stellarators \cite{4,5,6,7,8,9}. If the drift, bounce or transit frequencies of the energetic particles are resonant with the mode frequency, particle and diffusive losses increase, leading to a lower heating efficiency and more restrictive operational requirements for fusion ignition. In the case of resonance with plasma instabilities, internal kinks \cite{10,11} or ballooning modes \cite{12} can be kinetically destabilized. On the other hand, if the mode frequency is small, the interaction between background plasma and energetic particles may lead to a stabilizing effect (nonresonant limit) \cite{13,14}.

Super-Alfv\' enic alpha particles and energetic particles from neutral beam injection (NBI) can destabilize Alfv\' en Eigenmodes (AE), driven in the spectral gaps in the shear Alfv\' en continua \cite{15,16}. Periodic variations in the Alfv\' en speed produce frequency gaps associated with different Alfv\' en eigenmode families ($n$ is the toroidal mode and $m$ the poloidal mode), including: toroidicity induced Alfv\' en Eigenmodes (TAE) coupling $m$ with $m+1$ modes \cite{17,18}, helicity induced Alfv\' en Eigenmodes (HAE) coupling combinations of $n$ and $m$ modes \cite{19,20,21}, beta induced Alfv\' en Eigenmodes driven by compressibility effects (BAE) \cite{22,23}, Reversed-shear Alfv\' en Eigenmodes (RSAE) due to local maxima/minima in the rotational transform $\rlap{-} \iota$ profile \cite{24,25}, Global Alfv\' en Eigenmodes (GAE) observed in the minimum of the Alfv\' en continua \cite{26,27}, ellipticity induced Alfv\' en Eigenmodes (EAE) coupling $m$ with $m+2$ modes \cite{28,29}, noncircularity induced Alfv\' en Eigenmodes (NAE) coupling $m$ with $m+3$ or higher \cite{30,31}, mirror induced Alfv\' en Eigenmodes (MAE) coupling $n$ with $n+1$ at the same $m$ \cite{32}, as well as the kinetic version of the toroidicity induced Alfv\' en Eigenmodes (KTAE) \cite{33}. The presence of these modes in the plasma leads to larger alpha particle losses before thermalization \cite{34}, increasing the requirements for operations in self-sustained ignited plasmas, or a decrease of the NBI heating efficiency, also due to energetic particle losses \cite{35}.

TJ-II plasmas are heated by two neutral beam injectors (NBI), which inject hydrogen beams up to $32$ keV ($P_{NBI} = 0.5$ MW each). The hydrogen beams are 'co- /counter-' injected along/against the toroidal field leading to a small increase/decrease of the rotational transform by NBI driven currents. In addition, two gyrotrons heat the plasma, operating in the 2nd harmonic x-mode at 53.2 GHz ($P_{ECRH} = 0.3$ MW each) on and off axis. AE activity was measured in TJ-II operations \cite{21} as well as the effect of the rotational transform variation on the AE frequency \cite{36}. The aim of present study is to analyze the AE destabilization by energetic particles in TJ-II configurations, comparing simulation results and experimental observations. We also reproduce the AE frequency sweeping if the rotational transform profiles is displaced.

The analysis is performed using the FAR3D code \cite{37,38,39} supplemented by moment equations of the energetic ion density and parallel velocity that model the kinetic effects \cite{40,41}. The algorithm solves the reduced linear resistive MHD equations including the Landau damping/growth (linear wave-particle resonance effects) and geodesic acoustic waves (parallel momentum response of the thermal plasma) \cite{24}. The initial equilibrium is calculated by the VMEC code \cite{42} and the FAR3D code calculates the evolution of six field variables. A methodology has been developed to calibrate Landau-closure models against more complete kinetic models and optimize the closure coefficients \cite{24}. The model used in this paper includes Landau resonance couplings, but not fast ion FLR \cite{41} or Landau damping of the modes on the background ions/electrons \cite{40}. Methods for including these effects have been developed for the companion tokamak gyrofluid code TAEFL \cite{24}, and will be adapted to this 3D Landau fluid model as a topic for future research.​ The present model was already used to study the AE activity in LHD, indicating reasonable agreement with the observations \cite{43,44}.

This paper is organized as follows. The model equations, numerical scheme and equilibrium properties are described in section \ref{sec:model}. The simulation results are presented in section \ref{sec:simulation}. Finally, the conclusions of this paper are presented in section \ref{sec:conclusions}.

\section{Equations and numerical scheme \label{sec:model}}

For configurations with moderate $\beta$-values (of the order of the inverse aspect ratio) and high-aspect ratio, we use the method described in Ref.\cite{45} for the derivation of the reduced set of equations retaining the toroidal angle variation, to describe the evolution of the background plasma and fields. We obtain a reduced set of equations based upon the three-dimensional equilibrium including linear helical couplings between mode components. The moments of the kinetic equation truncated with a closure relation include the wave-particle resonnance effect of the energetic particle population effect. The Landau closure method was originally demonstrated for electrostatic modes \cite{46} and latter verified for electromagnetic energetic particle instabilities \cite{40}. By an appropriate choice of closure relations, this method incorporates the phase-mixing dynamics that leads to linear Landau damping and growth effects into fluid equations. These describe the evolution of the energetic particle density ($n_{f}$) and velocity moments parallel to the magnetic field lines ($v_{||f}$). The coefficients of the closure relation are selected to match a two-pole approximation of the plasma dispersion function. 

In the derivation of the reduced equations we assume high aspect ratio, medium $\beta$ (of the order of the inverse aspect ratio $\varepsilon=a/R_0$), small variation of the fields and small resistivity. The plasma velocity and perturbation of the magnetic field are defined as
\begin{equation}
 \mathbf{v} = \sqrt{g} R_0 \nabla \zeta \times \nabla \Phi, \quad\quad\quad  \mathbf{B} = R_0 \nabla \zeta \times \nabla \psi,
\end{equation}
where $\zeta$ is the toroidal angle, $\Phi$ is a stream function proportional to the electrostatic potential, and $\psi$ is the perturbation of the poloidal flux.

The equations, in dimensionless form, are

\begin{equation}
\label{eq:psieqls}
\frac{\partial \tilde \psi}{\partial t} =  \sqrt{g} B \nabla_\| \Phi  + \eta \varepsilon^2 J \tilde J^\zeta
\end{equation}

\begin{eqnarray} 
\label{eq:vorticityls}
\frac{{\partial \tilde U}}{{\partial t}} =  S^2 \left[{ \sqrt{g} B \nabla_\| J^\zeta - \frac{\beta_0}{2\varepsilon^2} \sqrt{g} \left( \nabla \sqrt{g} \times \nabla \tilde p \right)^\zeta }\right]   \nonumber\\
-  S^2 \left[{\frac{\beta_f}{2\varepsilon^2} \sqrt{g} \left( \nabla \sqrt{g} \times \nabla \tilde n_f \right)^\zeta }\right] 
\end{eqnarray} 

\begin{eqnarray}
\label{eq:pressure}
\frac{\partial \tilde p}{\partial t} = \frac{dp_{eq}}{d\rho}\frac{1}{\rho}\frac{\partial \tilde \Phi}{\partial \theta} +  \Gamma p_{eq}  \left[{ \sqrt{g} \left( \nabla \sqrt{g} \times \nabla \tilde \Phi \right)^\zeta - \nabla_\|  v_{\| th} }\right] 
\end{eqnarray} 

\begin{eqnarray}
\label{eq:vparth}
\frac{{\partial \tilde v_{\| th}}}{{\partial t}} =  -  \frac{S^2 \beta_0}{n_{0,th}} \nabla_\| p 
\end{eqnarray} 

\begin{eqnarray}
\label{eq:densf}
\frac{{\partial \tilde n_f}}{{\partial t}} =  -  \frac{S  v_{th,f}^2}{\omega_{cy}}\, \Omega_d (\tilde n_f) - S  n_{f0} \nabla_\| v_{\| f}  \nonumber\\
- \varepsilon^2  n_{f0} \, \Omega_d (\tilde \Phi) + \varepsilon^2 n_{f0} \, \Omega_* (\tilde  \Phi) 
\end{eqnarray} 

\begin{eqnarray}
\label{eq:vparf}
\frac{{\partial \tilde v_{\| f}}}{{\partial t}} =  -  \frac{S  v_{th,f}^2}{\omega_{cy}} \, \Omega_d (\tilde v_{\| f}) - \left( \frac{\pi}{2} \right)^{1/2} S  v_{th,f} \left| \nabla_\|  v_{\| f}  \right|  \nonumber\\
- \frac{S  v_{th,f}^2}{n_{f0}} \nabla_\| n_f + S \varepsilon^2  v_{th,f}^2 \, \Omega_* (\tilde \psi) 
\end{eqnarray} 

Here, $U =  \sqrt g \left[{ \nabla  \times \left( {\rho _m \sqrt g {\bf{v}}} \right) }\right]^\zeta$ is the vorticity and $\rho_m$ the ion and electron mass density. The toroidal current density $J^{\zeta}$ is defined as:
\begin{eqnarray}
J^{\zeta} =  \frac{1}{\rho}\frac{\partial}{\partial \rho} \left(-\frac{g_{\rho\theta}}{\sqrt{g}}\frac{\partial \psi}{\partial \theta} + \rho \frac{g_{\theta\theta}}{\sqrt{g}}\frac{\partial \psi}{\partial \rho} \right) \nonumber\\
- \frac{1}{\rho} \frac{\partial}{\partial \theta} \left( \frac{g_{\rho\rho}}{\sqrt{g}}\frac{1}{\rho}\frac{\partial \psi}{\partial \theta} + \rho \frac{g_{\rho \theta}}{\sqrt{g}}\frac{\partial \psi}{\partial \rho} \right)
\end{eqnarray}

The $v_{||th}$ is the parallel velocity of the thermal particles. The $n_{f}$ is normalized to the density at the magnetic axis $n_{f_{0}}$, $\Phi$ to $a^2B_{0}/\tau_{R}$ and $\Psi$ to $a^2B_{0}$. All lengths are normalized to a generalized minor radius $a$; the resistivity to $\eta_0$ (its value at the magnetic axis); the time to the resistive time $\tau_R = a^2 \mu_0 / \eta_0$; the magnetic field to $B_0$ (the averaged value at the magnetic axis); and the pressure to its equilibrium value at the magnetic axis. The Lundquist number $S$ is the ratio of the resistive time to the Alfv\' en time $\tau_{A0} = R_0 (\mu_0 \rho_m)^{1/2} / B_0$. $\rlap{-} \iota$ is the rotational transform, $v_{th,f} = \sqrt{T_{f}/m_{f}}$ the energetic particles thermal velocity normalized to the Alfv\' en velocity in the magnetic axis $v_{A0}$ and $\omega_{cy}$ the energetic particle cyclotron frequency times $\tau_{A0}$. The $q_{f}$ is the charge, $T_{f}$ the temperature and $m_{f}$ the mass of the energetic particles. $\Omega$ operators are defined as:

\begin{eqnarray}
\label{eq:omedrift}
\Omega_d = \frac{1}{2 B^4 \sqrt{g}}  \left[  \left( \frac{I}{\rho} \frac{\partial B^2}{\partial \zeta} - J \frac{1}{\rho} \frac{\partial B^2}{\partial \theta} \right) \frac{\partial}{\partial \rho}\right] \nonumber\\
-   \frac{1}{2 B^4 \sqrt{g}} \left[ \left( \rho \beta_* \frac{\partial B^2}{\partial \zeta} - J \frac{\partial B^2}{\partial \rho} \right) \frac{1}{\rho} \frac{\partial}{\partial \theta} \right] \nonumber\\ 
+ \frac{1}{2 B^4 \sqrt{g}} \left[ \left( \rho \beta_* \frac{1}{\rho} \frac{\partial B^2}{\partial \theta} -  \frac{I}{\rho} \frac{\partial B^2}{\partial \rho} \right) \frac{\partial}{\partial \zeta} \right]
\end{eqnarray}

\begin{eqnarray}
\label{eq:omestar}
\Omega_* = \frac{1}{B^2 \sqrt{g}} \frac{1}{n_{f0}} \frac{d n_{f0}}{d \rho} \left( \frac{I}{\rho} \frac{\partial}{\partial \zeta} - J \frac{1}{\rho} \frac{\partial}{\partial \theta} \right) 
\end{eqnarray}
Here the $\Omega_{d}$ operator is constructed to model the average drift velocity of a passing particle and $\Omega_{*}$ models its diamagnetic drift frequency. We also define the parallel gradient and curvature operators:
\begin{equation}
\label{eq:gradpar}
\nabla_\| f = \frac{1}{B \sqrt{g}} \left( \frac{\partial \tilde f}{\partial \zeta} +  \rlap{-} \iota \frac{\partial \tilde f}{\partial \theta} - \frac{\partial f_{eq}}{\partial \rho}  \frac{1}{\rho} \frac{\partial \tilde \psi}{\partial \theta} + \frac{1}{\rho} \frac{\partial f_{eq}}{\partial \theta} \frac{\partial \tilde \psi}{\partial \rho} \right)
\end{equation}
\begin{equation}
\label{eq:curv}
\sqrt{g} \left( \nabla \sqrt{g} \times \nabla \tilde f \right)^\zeta = \frac{\partial \sqrt{g} }{\partial \rho}  \frac{1}{\rho} \frac{\partial \tilde f}{\partial \theta} - \frac{1}{\rho} \frac{\partial \sqrt{g} }{\partial \theta} \frac{\partial \tilde f}{\partial \rho}
\end{equation}
with the Jacobian of the transformation:
\begin{equation}
\label{eq:Jac}
\frac{1}{\sqrt{g}} = \frac{B^2}{\varepsilon^2 (J+ \rlap{-} \iota I)}
\end{equation}

Equations~\ref{eq:pressure} and~\ref{eq:vparth} introduce the parallel momentum response of the thermal plasma, required for coupling to the geodesic acoustic waves, accounting the geodesic compressibility in the frequency range of the geodesic acoustic mode (GAM) \cite{48,49}.

The model uses the equilibrium flux coordinates $(\rho, \theta, \zeta)$ where $\rho$ is the generalized radial coordinate proportional to the square root of the toroidal flux function, and normalized to one at the edge, $\theta$ is the poloidal angle and $\zeta$ is the toroidal angle. The code also uses the Boozer coordinates \cite{47}. All functions have equilibrium and perturbation components represented as: $ A = A_{eq} + \tilde{A} $.

The FAR3D code uses finite differences in the radial direction and Fourier expansions in the two angular variables. The numerical scheme is semi-implicit in the linear terms.

\subsection{Equilibrium properties}

\begin{figure*}[h!]
\centering
\includegraphics[width=0.8\textwidth]{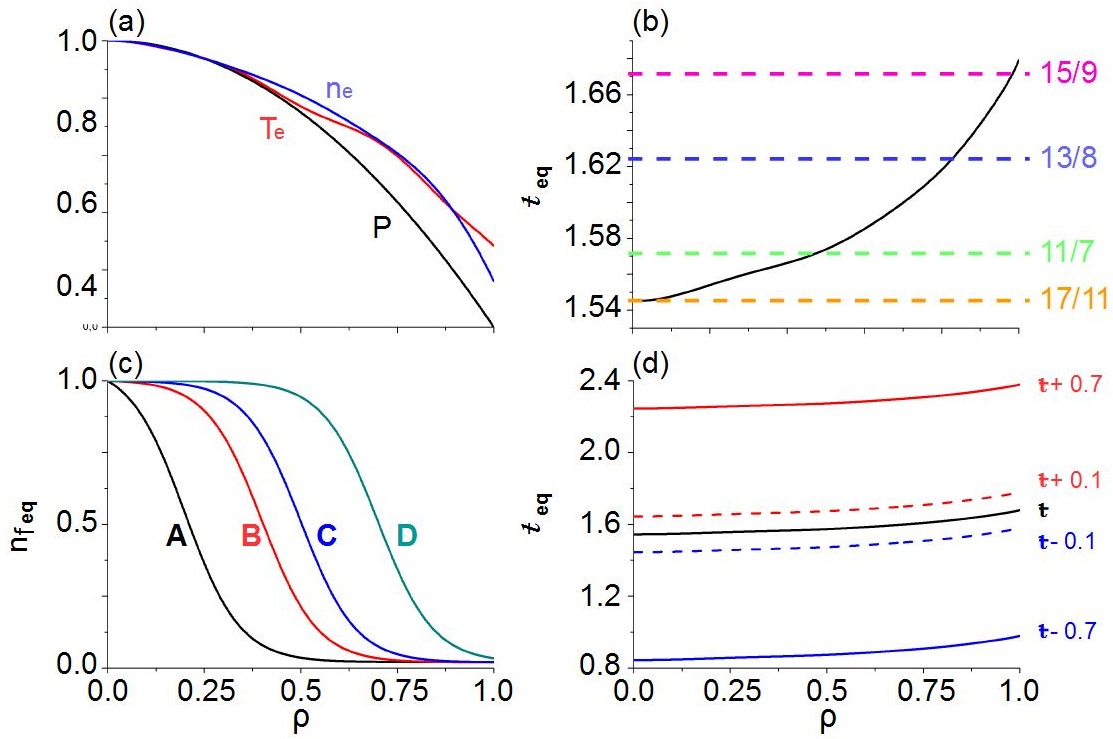}
\caption{(a) equilibrium pressure profile, (b) equilibrium rotational transform profile, (c) energetic particle density profiles and (d) equilibrium rotational transform profile displaced by $\Delta \rlap{-}\iota = \pm 0.7$ and $\pm 0.1$. Discharge number 18838.}\label{FIG:1}
\end{figure*}

We use a fixed boundary result from the VMEC equilibrium code \cite{42} calculated using the TJ-II reconstruction of discharges with AE activity \cite{21,36}.  
The electron density and temperature profiles were reconstructed by Thomson scattering data and electron cyclotron emission. The plasma is heated using two (co- and counter-) NBI injectors and two gyrotrons. The magnetic field at the magnetic axis is $1$ T, the averaged inverse aspect ratio $\varepsilon$ is $0.1285$ and $\beta_0$ is $0.0086$. The bulk electron density at the magnetic axis is $n_{e}(0) = 1.7 \cdot 10^{19}$ m$^{-3}$ and the bulk electron temperature at the magnetic axis is $T_{e}(0) = 0.6$ keV. The Alfv\' en velocity in the magnetic axis is $5.29 \cdot 10^{6}$ m/s. The energy of the injected particles by the NBI is $32$ keV ($v_{th,f} = 1.7 \cdot 10^{6}$ m/s and $v_{th,f}/v_{A0} = 0.32$) but we also consider in the study an averaged Maxwellian energy equal to the average energy of a slowing-down distribution with $32$ keV ($v_{th,f} = 1.11 \cdot 10^{6}$ m/s and $v_{th,f}/v_{A0}=0.21$). Figure~\ref{FIG:1} panel (a) shows the electron temperature, density and pressure equilibrium profiles. Panel (b) shows the equilibrium rotational transform profile including the resonant dominant modes. Panel (c) indicates the energetic particle density profiles, varying the location of the gradient: near the magnetic axis (case A), middle plasma (case B and C) and near the plasma periphery (case D). The energetic particle velocity profile is considered constant for simplicity. To analyze the effect of the rotational transform on the AE stability we perform simulations first displacing the equilibrium $\rlap{-}\iota$ profile by a coarse step $\Delta \rlap{-}\iota = 0.1$ between $\rlap{-}\iota = [-0.7,0.7]$ and then by a finer step $\Delta \rlap{-}\iota = 0.01$ between $\rlap{-}\iota = [-0.1,0.1]$, see Figure~\ref{FIG:1} panel (d).

\subsection{Simulations parameters}

The simulations are performed with a uniform radial grid of 1000 points. The dynamic and equilibrium toroidal/poloidal modes included in the study are summarized in table~\ref{Table:1} for the models with standard $\rlap{-}\iota$ profile ($\Delta \rlap{-}\iota = 0$). Table~\ref{Table:2} shows the extended mode selection in the simulations with the $\rlap{-}\iota$ profile displaced by $\Delta \rlap{-}\iota = 0.01$ and the further extended mode selection in the simulations with the $\rlap{-}\iota$ profile displaced by $\Delta \rlap{-}\iota = 0.1$.  In the following, the mode numbers used are referred to by $n/m$, consistent with the $\rlap{-}\iota$ definition.

\begin{table}[h]
\centering
\begin{tabular}{c | c }
Dyn. toroidal mode (n) & Poloidal mode (m) \\ \hline
$7$ & $[3,6]$ \\
$9$ & $[4,7]$ \\
$11$ & $[5,9]$ \\
$13$ & $[6,10]$ \\
$15$ & $[7,11]$ \\
$17$ & $[9,13]$ \\
Equil. toroidal mode (n) & Poloidal mode (m) \\ \hline
$0$ & $[0,5]$ \\
$4$ & $[0,5]$ \\
$8$ & $[2,7]$ \\
$12$ & $[5,10]$ \\
\end{tabular}
\caption{Dynamic and equilibrium toroidal and poloidal modes for the model with standard $\rlap{-}\iota$ profile ($\Delta \rlap{-}\iota = 0$)} \label{Table:1}
\end{table}

\begin{table}[h]
\centering
\begin{tabular}{c | c c}
Dyn. n & m ($\Delta \rlap{-}\iota = 0.01$) & m ($\Delta \rlap{-}\iota = 0.1$) \\ \hline
$7$ & $[2,6]$ & $[1,10]$ \\
$9$ & $[4,8]$ & $[2,12]$ \\
$11$ & $[5,9]$ & $[3,15]$ \\
$13$ & $[6,10]$ & $[4,17]$ \\
$15$ & $[7,12]$ & $[5,19]$ \\
$17$ & $[8,13]$ & $[6,22]$ \\
Equil. n & m ($\Delta \rlap{-}\iota = 0.01$) & m ($\Delta \rlap{-}\iota = 0.1$)  \\ \hline
$0$ & $[0,5]$ & $[0,5]$ \\
$4$ & $[0,5]$ & $[0,5]$ \\
$8$ & $[2,7]$ & $[1,9]$ \\
$12$ & $[5,10]$ & $[4,13]$ \\
\end{tabular}
\caption{Dynamic and equilibrium toroidal and poloidal modes for the simulations with the $\rlap{-}\iota$ profile displaced by $\Delta \rlap{-}\iota = 0.01$ and $0.1$} \label{Table:2}
\end{table}

The toroidal numbers $n=3$ ($n=5$) are not included in the analysis because the helical coupling with $n=7,11,15$ ($n=9,13,17$) helical family is weak. Even toroidal modes $n=4,8,12,16$ are only considered as equilibrium modes, not as dynamic modes, because the instabilities observed in TJ-II show mainly odd toroidal numbers. The kinetic closure moment equations (6) and (7) break the usual MHD parities. This is taken into account by including both parities $sin(m\theta + n\zeta)$ and $cos(m\theta + n\zeta)$ for all dynamic variables, and allowing for both a growth rate and real frequency in the eigenmode time series analysis. The convention of the code is, in case of the pressure eigenfunction, that $n > 0$ corresponds to the Fourier component $\cos(m\theta + n\zeta)$ and $n < 0$ to $\sin(-m\theta - n\zeta)$. For example, the Fourier component for mode $7/-2$ is $\cos(-2\theta + \zeta)$ and for the mode $-7/2$ is $\sin(-2\theta + \zeta)$. The magnetic Lundquist number is $S=5\cdot 10^6$ similar to the experimental value in the middle of the plasma.

The density ratio between energetic particles and bulk plasma ($n_{f}(0)/n_{e}(0)$) at the magnetic axis is controlled through the $\beta_{f}$ value. The ratio between energetic particle thermal velocity and Alfv\' en velocity in the magnetic axis ($v_{th,f}/v_{A0}$), controls the efficiency of the resonance coupling between AE and energetic particles. The cyclotron frequency is fixed at $\omega_{cy} = 27.1$ (normalized to the Alfv\' en time). 

\section{Simulation results \label{sec:simulation}}

The analysis is divided in two sections: simulations including toroidal mode couplings and simulations including helical couplings. We study the AE stability for different values of the $v_{th,f}/v_{A0}$ ratio (efficiency of the resonance coupling between AE and energetic particles), $\beta_{f}$ (energetic particle destabilization drive), $n_{f}$ profiles (effect of the location of the energetic particle density gradient) and the $\rlap{-}\iota$ profile (magnetic field topology).

\subsection{AE stability in TJ-II: toroidal couplings}     

In this section we study the AE stability for different values of the $v_{th,f}/v_{A0}$ ratio and $\beta_{f}$ values including only toroidal couplings (i.e., single n's, keeping only couplings across poloidal wave number). We also analyze the dependence of the pressure eigenfunction structure on the $v_{th,f}/v_{A0}$ ratio. 

\begin{figure*}[h!]
\centering
\includegraphics[width=0.8\textwidth]{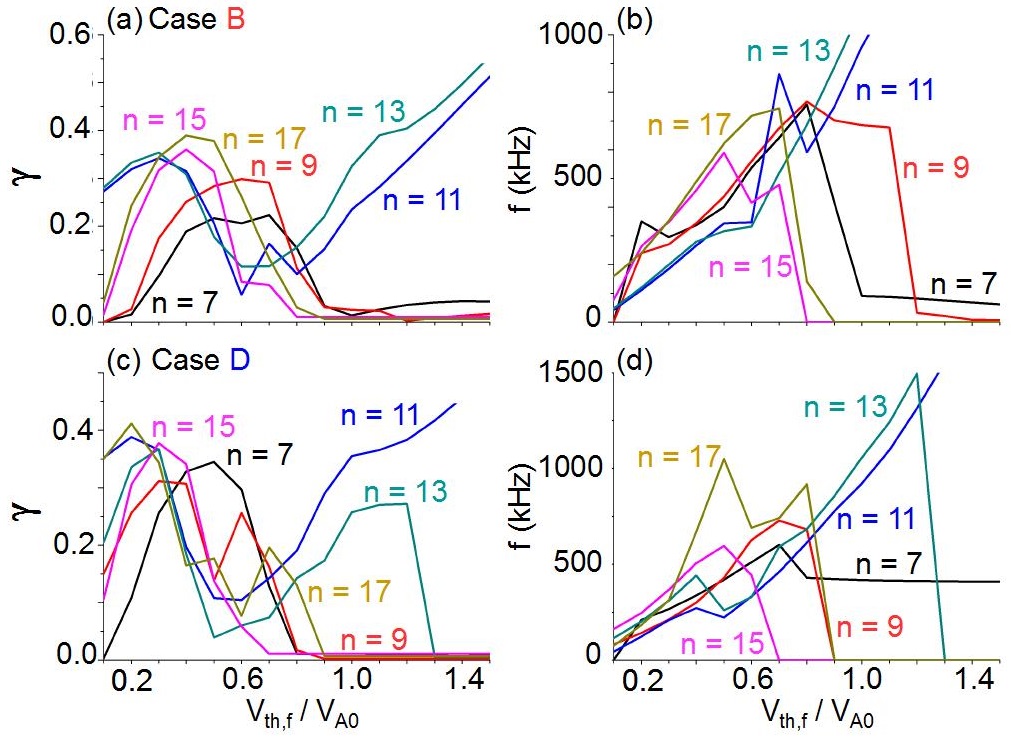}
\caption{Instability growth rate (a) and frequency (b) in the middle plasma (case B) and Instability growth rate (c) and frequency (d) in the middle plasma (case D) for different $v_{th,f}/v_{A0}$ ratios ($\beta_{f}=0.03$).}\label{FIG:2}
\end{figure*}

Figure~\ref{FIG:2} shows the instability growth rate ($\gamma$) and frequency (f) if the $n_{f}$ gradient is located in the middle of the plasma (Case B, panel a and b) or in the plasma periphery (Case D, panel c and d) for different $v_{th,f}/v_{A0}$ ratios (the TJ-II NBI operation regime is between $0.2-0.3$). If we compare the growth rate in cases B and D, the growth rate maxima are similar although they are displaced to a lower $v_{th,f}/v_{A0}$ ratio in case D, closer to the operational range of TJ-II NBI (between $0.2-0.3$). Consequently, for the parameter range of the TJ-II NBI the AE in the plasma periphery are easily destabilized. The transition from a AE to a standard MHD instability (growth rate and frequency show a large decrease) is observed for $n=9$ case B (case D) if $v_{th,f}/v_{A0} = 1.2$ ($0.9$), for $n=15$ case B (case D) if $v_{th,f}/v_{A0} = 0.8$ ($0.7$), for $n=13$ case D if $v_{th,f}/v_{A0} = 1.3$ and for $n=17$ in both cases if $v_{th,f}/v_{A0} = 0.9$. Modes $n=7$ and $n=11$ are AE unstable for all $v_{th,f}/v_{A0}$ ratios in the middle and plasma periphery. The frequency of the different toroidal modes in the TJ-II NBI operation regime is in the range of $[160,320]$ kHz for case B and $[130,280]$ kHz for case D. 

\begin{figure*}[h!]
\centering
\includegraphics[width=0.8\textwidth]{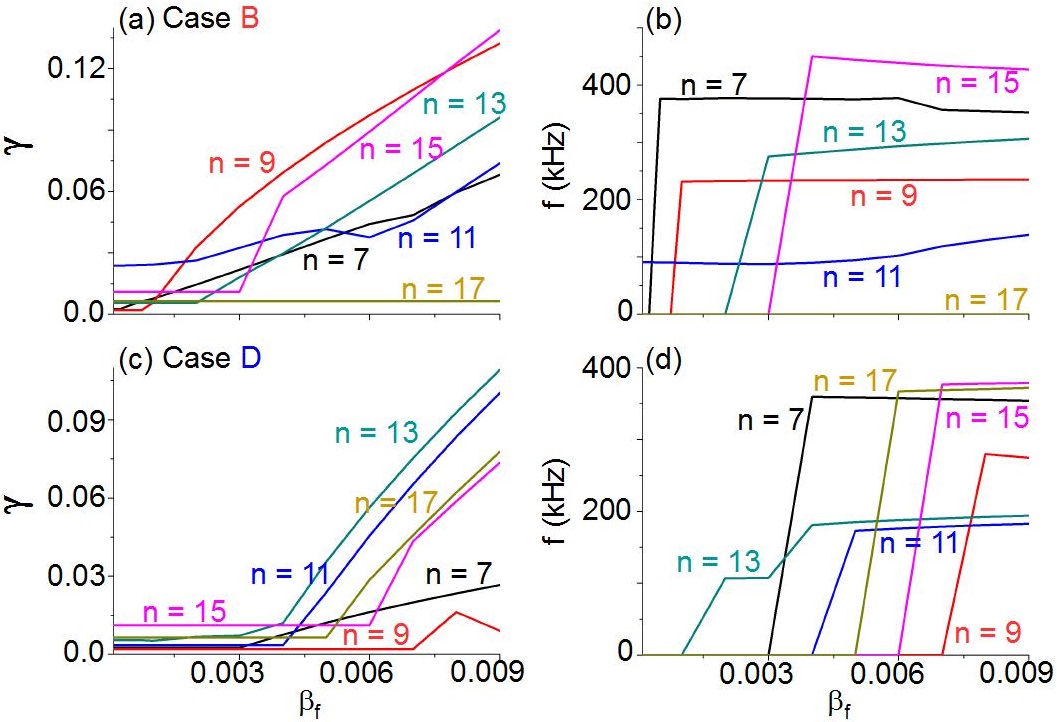}
\caption{Instability growth rate (a) and frequency (b) in the middle plasma (case B) and Instability growth rate (c) and frequency (d) in the middle plasma (case D) for different $\beta_{f}$ values ($v_{th,f}/v_{A0}=0.32$).}\label{FIG:3}
\end{figure*}

Figure~\ref{FIG:3} shows the instability growth rate and frequency if the $n_{f}$ gradient is located in the middle of the plasma (panel a and b) or in the plasma periphery (panel c and d) for different $\beta_{f}$ values (the TJ-II NBI operation regime is between $0.002-0.004$). The $\beta_{f}$ threshold to destabilize AEs is lower in the middle plasma than in the plasma periphery. The $n=17$ mode is AE stable in the middle plasma although $n=11$ mode is AE unstable if $\beta_{f} > 0.0001$. The AE frequency is in the range of $[100,440]$ kHz in the middle plasma and between $[180,390]$ kHz in the plasma periphery. The $n=9$ AE is dominant if $\beta_{f} > 0.001$, $n=11$ AE if $\beta_{f} < 0.001$ and $n=15$ AE if $\beta_{f} > 0.007$ in the middle plasma. The $n=9$ AE frequency is $240$ kHz, $n=11$ AE frequency is $90$ kHz (increasing to $140$ kHz for large $\beta_{f}$ values) and $n=15$ AE frequency is $440$ kHz. In the plasma periphery, if $\beta_{f} < 0.003$ the plasma is stable to AE instabilities and unstable to $n=15$ ballooning modes. If $\beta_{f} > 0.003$, $n=13$ and $n=11$ AE dominates with a frequency of $190$ and $180$ kHz respectively.

\begin{figure*}[h!]
\centering
\includegraphics[width=0.8\textwidth]{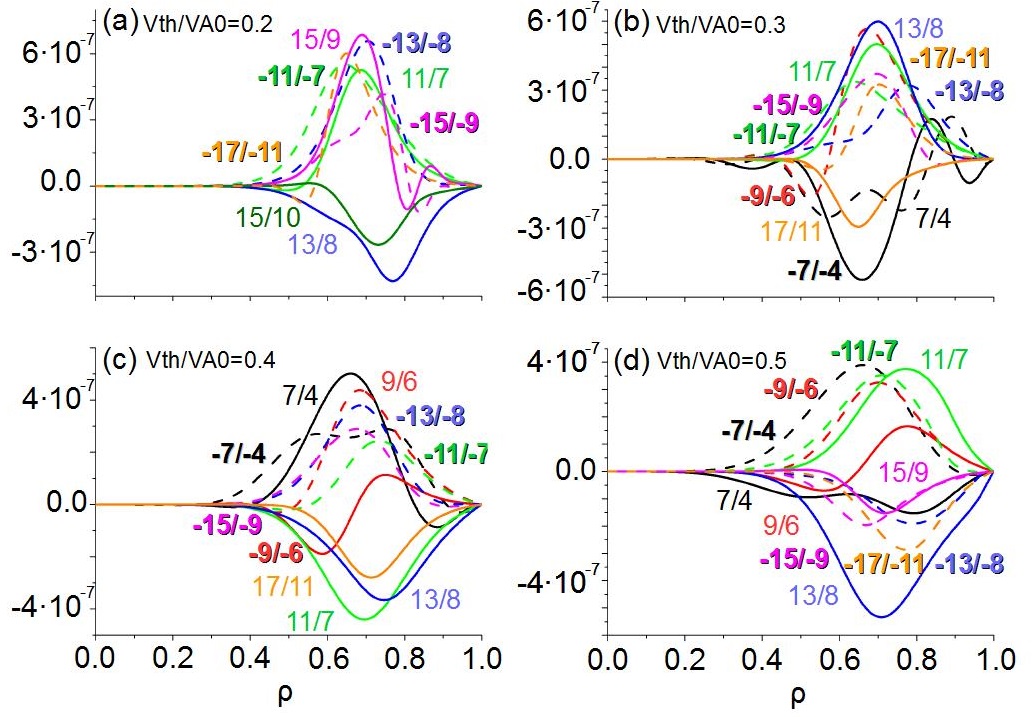}
\caption{Pressure eigenfunctions of the dominant poloidal mode for each toroidal mode if $v_{th,f}/v_{A0} = 0.2$ (panel a), $0.3$ (panel b), $0.4$ (panel c) and $0.5$ (panel d) in the plasma periphery. Modes with negative toroidal number (dotted line and bold numbers) and modes with positive toroidal number (solid line and thin numbers). If $n > 0$ the Fourier component is $\cos(m\theta + n\zeta)$ and if $n < 0$ it is $\sin(-m\theta - n\zeta)$.}\label{FIG:4}
\end{figure*}

Figure~\ref{FIG:4} shows the pressure eigenfunctions of the dominant poloidal component for each toroidal mode if $v_{th,f}/v_{A0} = 0.2$ (panel a), $0.3$ (panel b), $0.4$ (panel c) and $0.5$ (panel d) in the plasma periphery. The cases with $v_{th,f}/v_{A0} = 0.2$ and $0.3$ show a large correlation between local peaks in the eigenfunctions at different toroidal mode numbers. This characteristic implies that helical couplings will likely be important; these cases will be re-examined in Section 3.2 including helical couplings. The eigenfunction width is larger as the $v_{th,f}/v_{A0}$ ratio increases and the correlation between local peaks of different toroidal modes is smaller. The cases $v_{th,f}/v_{A0} = 0.4$ and $0.5$ show weak helical couplings, although the toroidal couplings can lead to the destabilization of TAEs, for example $n=11$ and $n=13$ TAE in $v_{th,f}/v_{A0} = 0.5$ simulations with frequencies of $345$ and $320$ kHz.

\begin{figure*}[h!]
\centering
\includegraphics[width=0.8\textwidth]{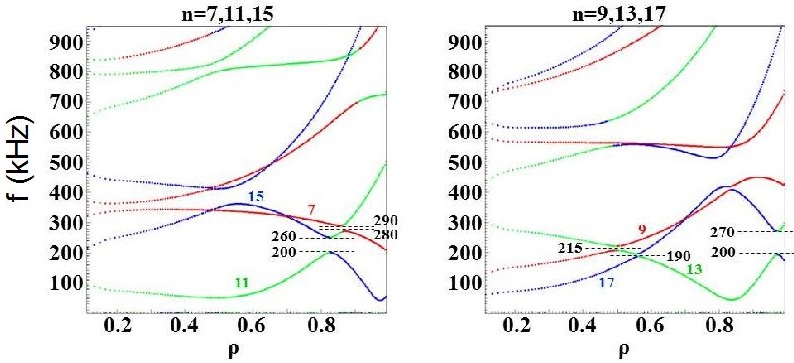}
\caption{Alfv\' en gaps for the standard $\rlap{-}\iota$ profile}\label{FIG:5}
\end{figure*}

Figure~\ref{FIG:5} shows the Alfv\' en continuum gap structure of the TJ-II equilibria including helical couplings \cite{21}.The Alfv\' en gaps indicate the potential for modes of the $n=9-13$ AE with $f=215$ kHz and the $n=13-17$ AE with $f=190$ kHz in the middle plasma. In the plasma periphery, the $n=11-15$ AE with $f=200-260$ kHz, the $n=7-11$ AE with $f=280-290$ kHz and the $n=13-17$ AE with $f=200-270$ kHz can be destabilized. On the other hand, simulations with toroidal couplings predict an unstable $n=9$ AE with $f=240$ kHz in the middle of the plasma and a marginally unstable $n=13$ energetic particle mode (EPM) in the periphery for the TJ-II NBI operational regime. The pressure eigenfunctions of $v_{th,f}/v_{A0} = 0.2$ and $0.3$ cases suggest large helical couplings in the TJ-II NBI operational regime, so a model that only includes toroidal couplings will not reproduce TJ-II observations, particularly in the plasma periphery. In the next section we extend the study to include the effect of the helical couplings.

\subsection{AE stability in TJ-II: helical coupling}     

In this section we study the AE stability including helical couplings, comparing simulation results and TJ-II observations for different values of the $v_{th,f}/v_{A0}$ ratio and $\beta_{f}$ values \cite{21}. 

\begin{figure*}[h!]
\centering
\includegraphics[width=0.8\textwidth]{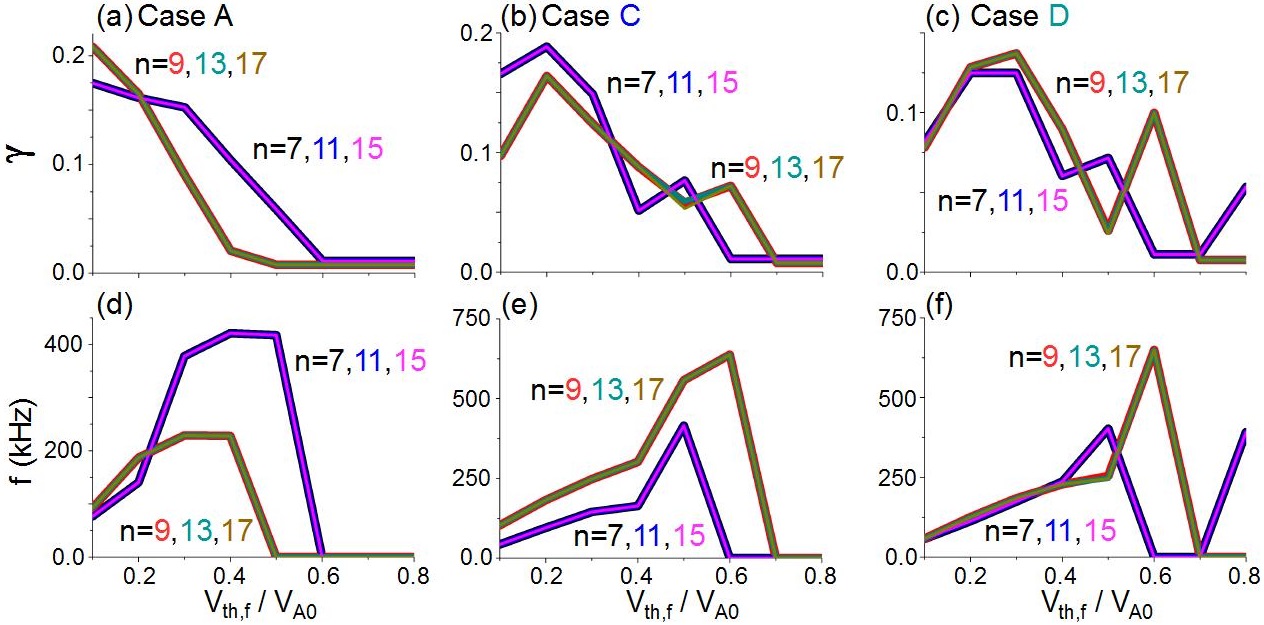}
\caption{Instability growth rate (a) and frequency (d) near the magnetic axis (case A), instability growth rate (b) and frequency (e) in the middle plasma (case C) and instability growth rate (c) and frequency (f) in the plasma periphery for different $v_{th,f}/v_{A0}$ ratios ($\beta_{f}=0.01$).}\label{FIG:6}
\end{figure*}

Figure~\ref{FIG:6} shows the growth rate and frequency for different $n_{f}$ profiles and $v_{th,f}/v_{A0}$ ratios, fixed $\beta_{f} = 0.01$. The toroidal mode numbers $n=7,11,15$ and $n=9,13,17$ are helically coupled and evolve together. The TJ-II NBI operational regime is in the range of $v_{th,f}/v_{A0}$ ratios with the largest growth rates for all the $n_{f}$ profiles, although the local maxima of the frequency is displaced to higher $v_{th,f}/v_{A0}$ ratios, between $0.5$ and $0.6$. In the TJ-II NBI operational regime the helical families show similar growth rates, so both AEs can be destabilized, except if the $n_{f}$ profiles is near the magnetic axis and $v_{th,f}/v_{A0} = 0.3$, leading to a dominant $n=7,11,15$ AE. The $n=7,11,15$ AE frequency near the magnetic axis is in the range of the $[130,350]$ kHz, $[110,150]$ kHz in the middle plasma and $[100,160]$ kHz in the plasma periphery. The $n=9,13,17$ AE frequency near the magnetic axis is in the range of the $[190,220]$ kHz, $[200,250]$ kHz in the middle plasma and $[110,170]$ kHz in the plasma periphery. For a ratio $v_{th,f}/v_{A0} < 0.2$, both helical families show similar growth rates and frequencies, smaller compared to the TJ-II NBI operational regime, except near the magnetic axis where $n=9,13,17$ AE is dominant. Increasing the $v_{th,f}/v_{A0}$ ratio further from the TJ-II NBI operational regime leads to the AE stabilization in the plasma core if $v_{th,f}/v_{A0} > 0.4$ for $n=9,13,17$ AE, and $n=7,11,15$ AE if $v_{th,f}/v_{A0} > 0.5$. In the middle of the plasma, if $v_{th,f}/v_{A0} > 0.5$ $n=9,13,17$ AE is stable as well as the $n=7,11,15$ AE if $v_{th,f}/v_{A0} > 0.6$. In the outer plasma, the $n=9,13,17$ AE is stable if $v_{th,f}/v_{A0} > 0.5$ and the $n=7,11,15$ AE is stable if $v_{th,f}/v_{A0} > 0.6$.

\begin{figure*}[h!]
\centering
\includegraphics[width=0.8\textwidth]{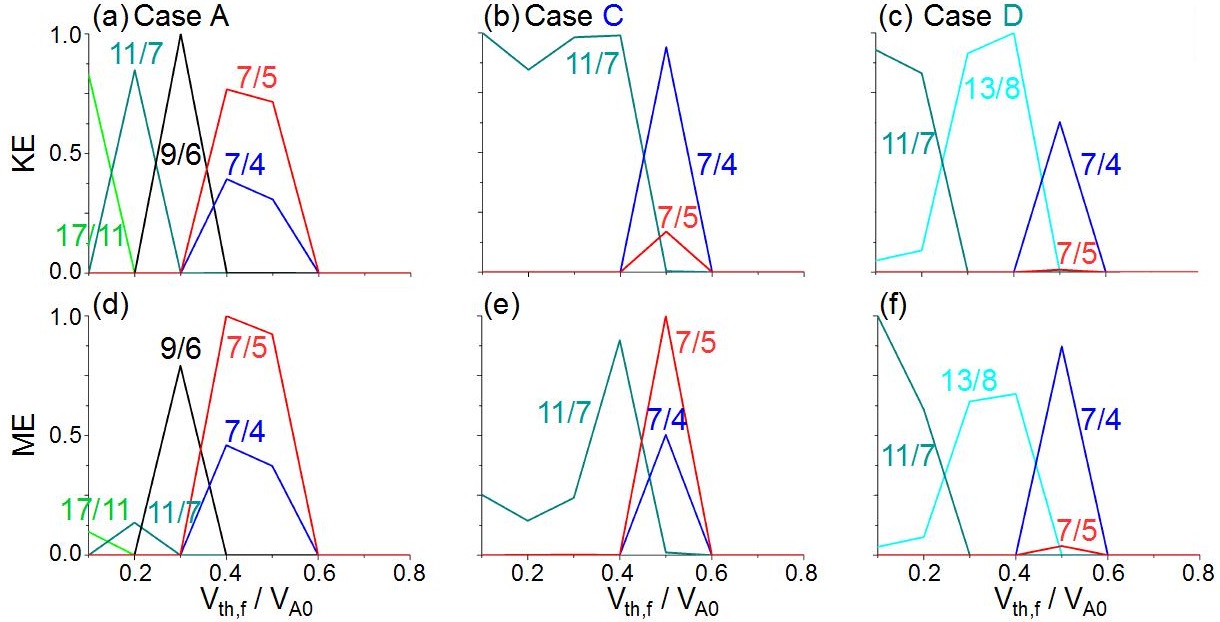}
\caption{KE and ME of the dominant modes for case A (panels a and d), case C (panels b and e) and case D (panels c and f) for different $v_{th,f}/v_{A0}$ ratios ($\beta_{f}=0.01$).}\label{FIG:7}
\end{figure*}

Figure~\ref{FIG:7} shows the normalized kinetic energy (KE) and magnetic energy (ME) of the dominant modes for different $n_{f}$ profiles and $v_{th,f}/v_{A0}$ ratios, at fixed $\beta_{f} = 0.01$. In the TJ-II NBI operational regime, the modes $11/7$ and $9/6$ are dominant in the plasma core, $11/7$ is dominant in the middle plasma, and $11/7$ and $13/8$ are present in the plasma periphery. Below the TJ-II NBI operational regime, the $11/7$ mode dominates in all the plasma if $v_{th,f}/v_{A0}=0.2$ and the AE frequency is $f \approx 130$ kHz (case A) and $75$ kHz (case C), a frequency below a minima of the continuum plot (see Fig. 5). Consequently the AE observed in case A and C is an extremal mode ($df/d\rho = 0$) so it can be identified as a GAE. In case D, the instability frequency is $f \approx 100$ kHz but ($df/d\rho \neq 0$) so it cant be identified as a GAE but rather seemed as a BAE. If $v_{th,f}/v_{A0}=0.1$, the $11/7$ mode dominates in the middle and outer plasma and the $17/11$ mode in the core. These instabilities have a lower frequency than those with other $v_{th,f}/v_{A0}$ ratios, pointing out the destabilization of EPM. If $v_{th,f}/v_{A0}>0.2$, above the TJ-II NBI operational regime, the modes $7/5$ and $7/4$ are dominant in the whole plasma, implying the destabilization of an $n=7$ TAE.

\begin{figure*}[h!]
\centering
\includegraphics[width=0.8\textwidth]{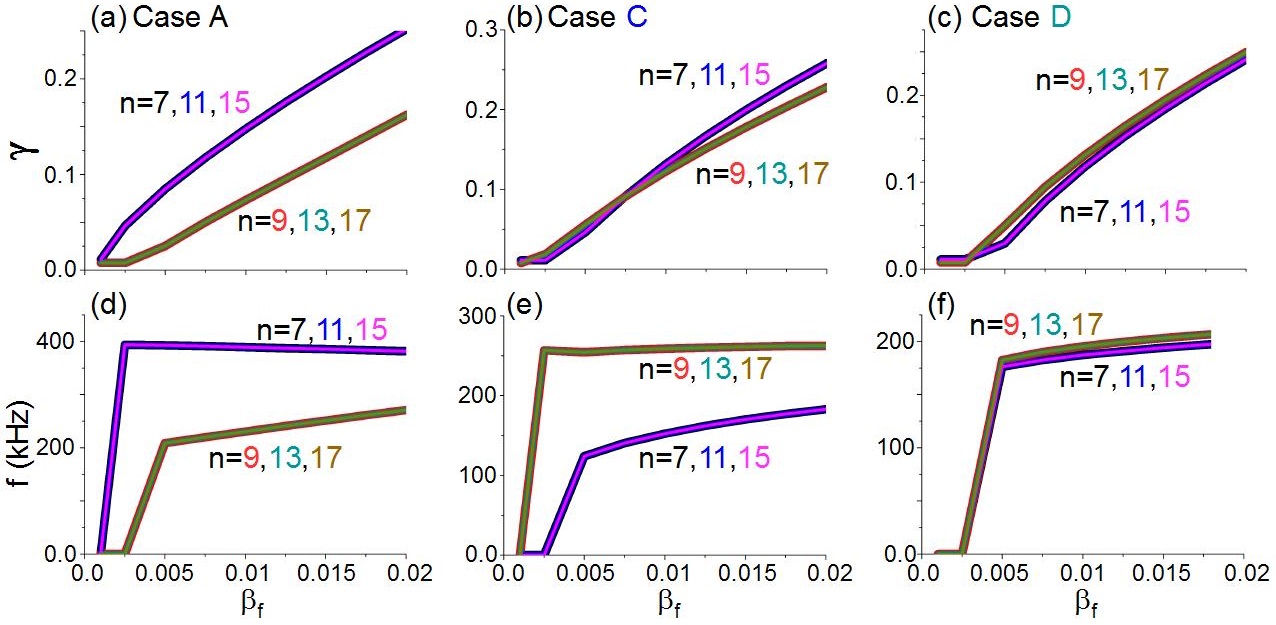}
\caption{Instability growth rate (a) and frequency (d) near the magnetic axis (case A), instability growth rate (b) and frequency (e) in the middle plasma (case C) and instability growth rate (c) and frequency (f) in the plasma periphery for different $\beta_{f}$ values ($v_{th,f}/v_{A0}=0.32$).}\label{FIG:8}
\end{figure*}

Figure~\ref{FIG:8} shows the growth rate and frequency for different $n_{f}$ profiles and $\beta_{f}$ values, for a fixed $v_{th,f}/v_{A0} = 0.32$. Near the magnetic axis, the $n=7,11,15$ HAE is dominant with a $\beta_{f}$ threshold smaller than $0.0025$, while the $n=9,13,17$ $\beta_{f}$ threshold is between $0.0025$ and $0.005$. The $n=7,11,15$ HAE frequency is $400$ kHz and $n=9,13,17$ frequency is in between $210$ kHz and $275$ kHz. The $n=9,13,17$ instability shows a large frequency variation for different $\beta_{f}$ values so it should be identified as an energetic particle mode (EPM), not a HAE. In the middle region of the plasma, the growth rate of both helical mode families is similar but the $\beta_{f}$ threshold is less than $0.0025$ for $n=9,13,17$ HAE and between $0.0025$-$0.005$ for $n=7,11,15$ EPM. For large $\beta_{f}$ values  the $n=7,11,15$ EPM is dominant. The $n=9,13,17$ HAE frequency is $250$ kHz and the $n=7,11,15$ EPM frequency is in the range of $[130,190]$ kHz. In the plasma periphery both helical mode families show similar growth rates and the $\beta_{f}$ threshold is less than $0.0025$. Both instabilities are HAEs with a frequency of $200$ kHz. 

\begin{figure*}[h!]
\centering
\includegraphics[width=0.8\textwidth]{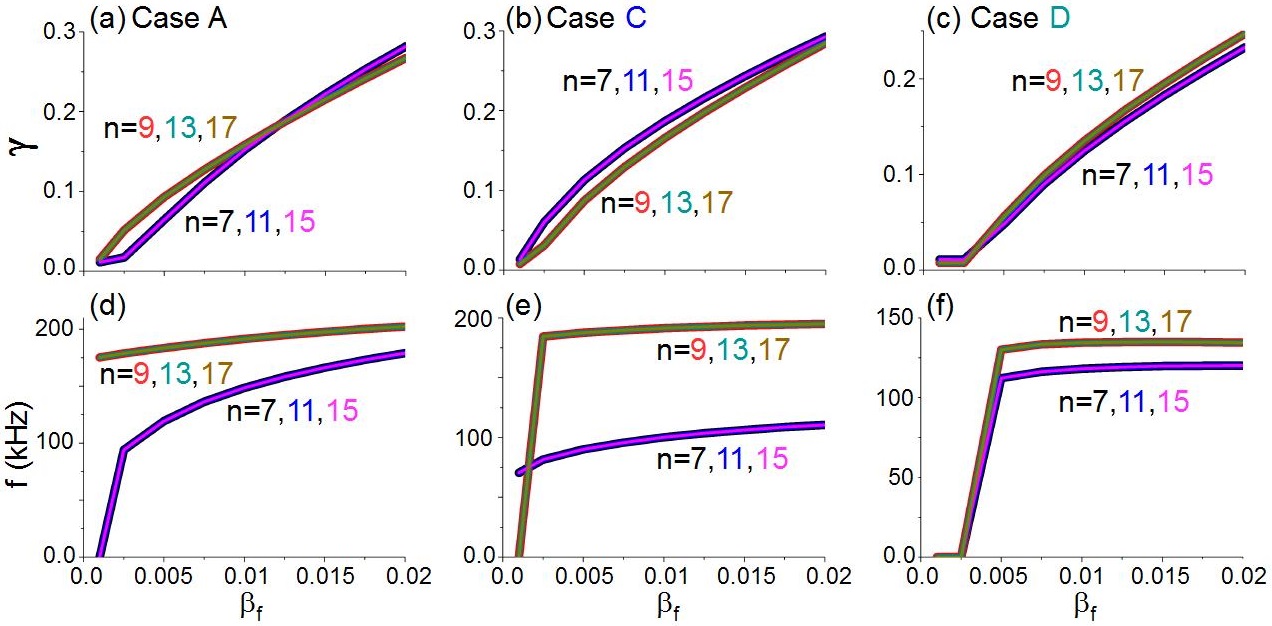}
\caption{Instability growth rate (a) and frequency (d) near the magnetic axis (case A), instability growth rate (b) and frequency (e) in the middle plasma (case C) and instability growth rate (c) and frequency (f) in the plasma periphery for different $\beta_{f}$ values ($v_{th,f}/v_{A0}=0.21$).}\label{FIG:9}
\end{figure*}

To improve the comparison between simulations and TJ-II AE measurements, we perform a new set of simulations, reducing the $v_{th,f}/v_{A0}$ ratio from $0.32$ to $0.21$, to model the effects of a slowing-down distribution on the average energy of the energetic particles. Figure~\ref{FIG:9} shows the results of such a study for different $n_{f}$ profiles and $\beta_{f}$ values. Near the magnetic axis, an $n=9,13,17$ EPM with a frequency between $175-200$ kHz and a $\beta_{f}$ threshold smaller than $0.001$ dominates. For larger $\beta_{f}$ values, more than $0.01$, an $n=7,11,15$ EPM dominates with a frequency between $[100,175]$ kHz and a $\beta_{f}$ threshold of $0.0025$. In the middle of the plasma, the $n=7,11,15$ EPM is dominant with a $\beta_{f}$ threshold smaller than $0.001$ and a frequency between $[70,120]$ kHz. The $\beta_{f}$ threshold of the $n=9,13,17$ HAE with $f=185$ kHz is smaller than $0.0025$. In the plasma periphery both $n=7,11,15$ and $n=9,13,17$ HAE show similar growth rates (the $n=9,13,17$ HAE is slightly larger for high $\beta_{f}$ values) with a $\beta_{f}$ threshold between $0.0025$ and $0.005$. The $n=7,11,15$ HAE frequency is $120$ kHz and the $n=9,13,17$ HAE frequency is $135$ kHz. 

\begin{figure*}[h!]
\centering
\includegraphics[width=0.8\textwidth]{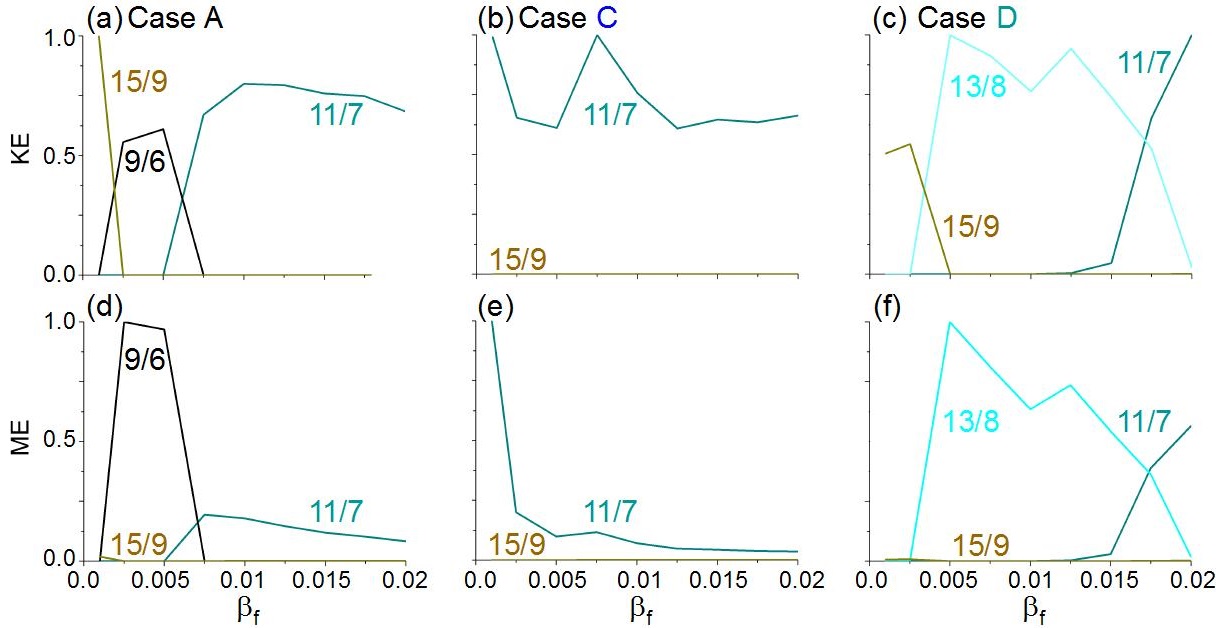}
\caption{KE and ME of the dominant modes for case A (panels a and d), case C (panels b and e) and case D (panels c and f) for different $\beta_{f}$ values  ($v_{th,f}/v_{A0}=0.21$).}\label{FIG:10}
\end{figure*}

Figure~\ref{FIG:10} shows the normalized kinetic energy (KE) and magnetic energy (ME) of the dominant modes for different $n_{f}$ profiles and $\beta_{f}$ values, at fixed $v_{th,f}/v_{A0}=0.21$. For $\beta_{f}$ values below the threshold, a $15/9$ ballooning mode is destabilized in the core and plasma periphery, although the EPM $11/7$ is dominant in the middle of the plasma. For the $\beta_{f}$ values in the TJ-II NBI operational regime, the $9/6$ mode dominates in the plasma core, $11/7$ in the middle of the plasma and $13/8$ in the periphery, coherent with the destabilization of an HAE. For $\beta_{f}$ values above the TJ-II NBI operation regime, the mode $11/7$ dominates in the plasma core and middle of the plasma, also in the plasma periphery if $\beta_{f} > 0.0175$, pointing out the destabilization of a $11/7$ GAE (also measured in TJ-II \cite{50}).

In summary, in the NBI TJ-II operational regime ($\beta_{f} =0.0025-0.005$), $n=9,13,17$ HAEs with $f=180$ kHz are unstable in the plasma core even if the NBI destabilization is weak ($\beta_{f} < 0.001$). In the middle of the plasma an $n=7,11,15$ EPM can be destabilized even if the destabilization driven by the NBI is weak, followed by an $n=9,13,17$ HAE with $f=185$ kHz if $\beta_{f} > 0.001$. In the plasma periphery, an $n=7,11,15$ HAE at $120$ kHz and an $n=9,13,17$ HAE at $135$ kHz are destabilized if $\beta_{f} > 0.0025$. These results are comparable to TJ-II measurements of AE activity with frequency between $[50,300]$ kHz \cite{21}. The results are also coherent with the Alfv\' en gap frequencies in the middle and outer plasma.

\subsubsection{AE stabilization by optimized $\rlap{-}\iota$ profiles in TJ-II}     

In this section we analyze the effect of the $\rlap{-}\iota$ profile, namely the magnetic field topology, on AE stability. We perform a set of simulations displacing the standard $\rlap{-}\iota$ profile by $\rlap{-}\iota \pm \Delta\rlap{-}\iota\cdot i$ with $\Delta\rlap{-}\iota = 0.1$ and $i=[1,7]$.

\begin{figure*}[h!]
\centering
\includegraphics[width=0.8\textwidth]{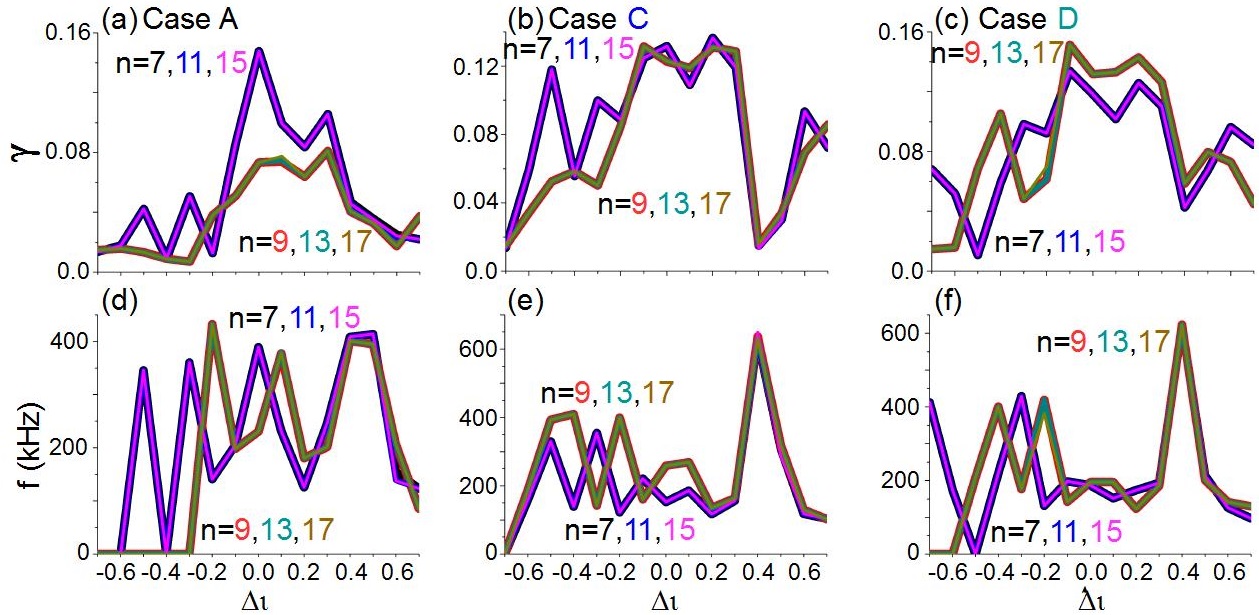}
\caption{Growth rate and frequency of the dominant modes for case A (panels a and d), case C (panels b and e) and case D (panels c and f) if $\rlap{-}\iota$ profile is displaced by $\rlap{-}\iota \pm \Delta\rlap{-}\iota\cdot i$ if $\Delta\rlap{-}\iota = 0.1$ and $i=[1,7]$ ($v_{th,f}/v_{A0}=0.32$ and $\beta_{f}=0.01$).}\label{FIG:11}
\end{figure*}

Figure~\ref{FIG:11} shows the growth rate and frequency for different $n_{f}$ profiles and $\Delta \rlap{-}\iota$, keeping a constant $v_{th,f}/v_{A0}=0.32$ and $\beta_{f}=0.01$. The configuration with the standard $\rlap{-}\iota$ profile and displacements of $\Delta\rlap{-}\iota < 0.4$ lead to the largest growth rates, particularly in the middle and outer plasma regions. If the $\rlap{-}\iota$ profile is displaced by a positive $\Delta\rlap{-}\iota > 0.4$, the AE growth rate and frequency decreases, except if $\Delta\rlap{-}\iota = 0.4$, showing a local minima of the growth rate and a local maxima of the frequency. If the $\rlap{-}\iota$ profile is displaced by a negative $\Delta\rlap{-}\iota$, the growth rate decreases and the frequency increases. The $n=9,13,17$ AEs are stable in the plasma core if $\Delta\rlap{-}\iota < -0.2$, in the middle plasma if $\Delta\rlap{-}\iota < -0.6$ and in the plasma periphery if $\Delta\rlap{-}\iota < -0.5$. The growth rate of the $n=7,11,15$ AE also decreases compared to the standard $\rlap{-}\iota$ profile configuration, although it is never stabilized in the whole plasma for any $\Delta\rlap{-}\iota$ displacement. Consequently, the most efficient configuration for reducing AE instability in the plasma core and middle region of the plasma requires an $\rlap{-}\iota = [0.845,0.979]$. No optimal $\rlap{-}\iota$ profile exists for the plasma periphery to fully stabilize AEs, although if $\rlap{-}\iota = [0.945,1.079]$ the $n=9,13,17$ AEs are stable and $n=7,11,15$ AE growth rate is 4 times smaller compared with the standard $\rlap{-}\iota$ profile configuration.  

This study was performed using the same VMEC equilibria, so the displacement of the $\rlap{-}\iota$ profile can lead to inconsistencies, although such discrepancies don't invalidate the trends of the analysis. 

In the next section we will perform a similar study, but reducing the $\rlap{-}\iota$ profile displacement to $\Delta\rlap{-}\iota = 0.01$, thus avoiding large excursions from the original equlibria and improving the reliability of the analysis.

\subsubsection{Effect of the $\rlap{-}\iota$ profile on TJ-II AE activity}     

The aim of this section is to reproduce the TJ-II observations for discharges with time-varying $\rlap{-}\iota$ profiles \cite{36}. Such TJ-II operations show large AE frequency sweeps, leading to a sawtooth-like evolution of the AEs signal as the $\rlap{-}\iota$ profiles evolves, similar to Alfv\' en cascades in reversed-shear tokamak plasmas \cite{24,51}.

\begin{figure*}[h!]
\centering
\includegraphics[width=0.8\textwidth]{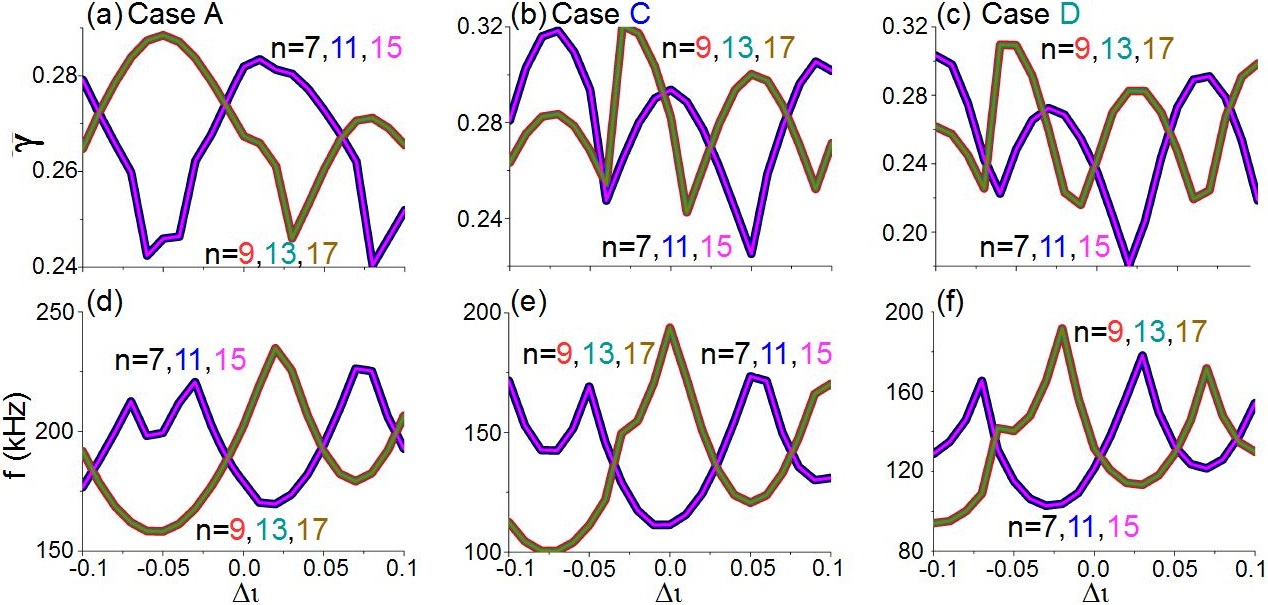}
\caption{Growth rate and frequency of the dominant modes for case A (panels a and d), case C (panels b and e) and case D (panels c and f) if $\rlap{-}\iota$ profile is displaced by $\rlap{-}\iota \pm \Delta\rlap{-}\iota\cdot i$ if $\Delta\rlap{-}\iota = 0.01$ and $i=[1,10]$ ($v_{th,f}/v_{A0}=0.21$ and $\beta_{f}=0.02$).}\label{FIG:12}
\end{figure*}

\begin{figure*}[h!]
\centering
\includegraphics[width=0.8\textwidth]{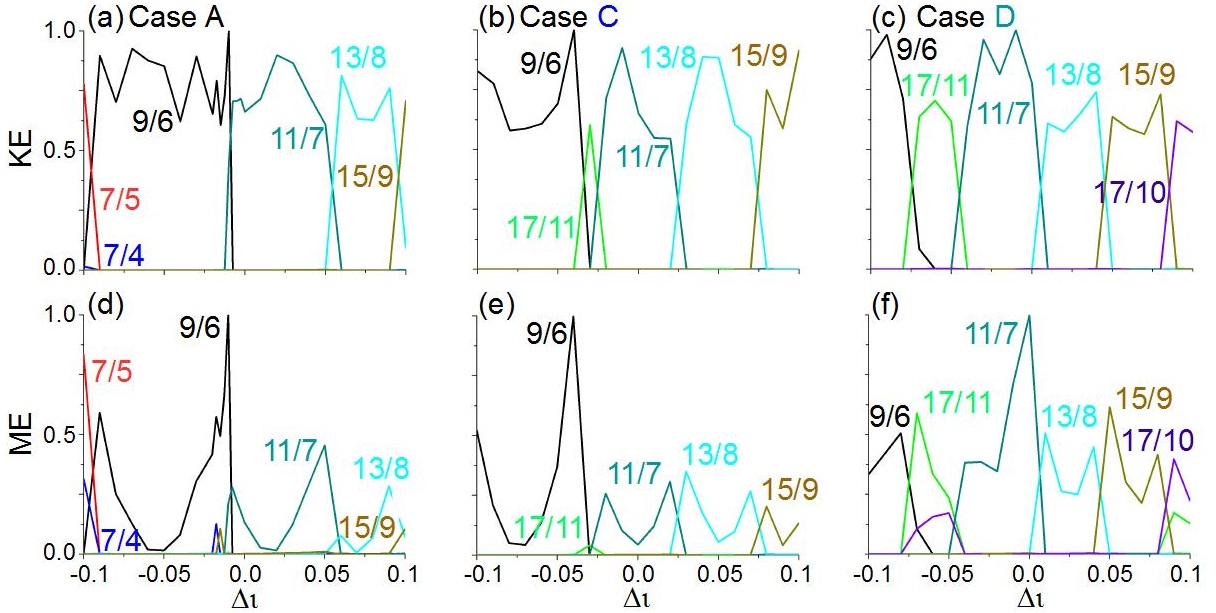}
\caption{KE and ME of the dominant modes for case A (panels a and d), case C (panels b and e) and case D (panels c and f) if $\rlap{-}\iota$ profile is displaced by $\rlap{-}\iota \pm \Delta\rlap{-}\iota\cdot i$ if $\Delta\rlap{-}\iota = 0.01$ and $i=[1,10]$ ($v_{th,f}/v_{A0}=0.21$ and $\beta_{f}=0.02$).}\label{FIG:13}
\end{figure*}

Figure~\ref{FIG:12} shows the growth rate and frequency for different $n_{f}$ profiles and $\rlap{-}\iota$ profiles displaced by $\rlap{-}\iota \pm \Delta\rlap{-}\iota\cdot i$ with $\Delta\rlap{-}\iota = 0.01$ and $i=[1,10]$, fixing $v_{th,f}/v_{A0} = 0.21$ and $\beta_{f}=0.02$. There is a sawtooth-like evolution of both helical mode families growth rate and frequency as the $\rlap{-}\iota$ profiles are displaced, showing inverse correlations between local growth rate and frequency minima/maxima. Furthermore, the local maxima of the growth rate of $n=7,11,15$ helical family are correlated with a local growth rate minima of $n=9,13,17$ (except in the middle plasma if $\Delta\rlap{-}\iota < -0.05$, showing both helical families a local growth rate maxima and frequency minima). The growth rate increases (decreases) about 20 $\%$ between local maxima and minima. The frequency oscillates between $[160,230]$ kHz in the inner plasma and between $[95,195]$ kHz in the middle and outer plasma. Figure~\ref{FIG:13} shows the KE and ME of the dominant modes as the $\rlap{-}\iota$ profile is displaced, pointing out that the frequency sweeping evolution is driven by the change of the instability dominant mode, associated to different helical families. For example, in the inner plasma, if we apply a negative displacement of $\Delta \rlap{-}\iota = -0.01$, there is a transition between the dominant mode $11/7$ ($n=7,11,15$ helical family) for the standard $\rlap{-}\iota$ configuration, to a dominant $9/6$ mode ($n=9,13,17$ helical family). If we keep displacing the $\rlap{-}\iota$ profile up to $\Delta \rlap{-}\iota = -0.05$, the AE frequency decreases from $190$ kHz to $160$ kHz. If we apply a positive $\rlap{-}\iota$ profile displacement between the standard case and $\Delta\rlap{-}\iota = 0.03$, the mode $11/7$ is dominant and the AE frequency decreases from $190$ kHz to $170$ kHz. If the $\rlap{-}\iota$ profile is further displaced to $\Delta\rlap{-}\iota > 0.07$, the $n=9,13,17$ AE is dominant again, particularly the $13/8$ mode. The same behavior is observed in the middle and outer plasma, although new modes are destabilized, as the $17/11$ mode in the middle plasma or the $17/10$ mode in the outer plasma, leading to a faster evolution of the growth rate and frequency with $\Delta\rlap{-}\iota$ displacement. In case A, we increase the resolution of the $\rlap{-}\iota$ profile displacement to $\Delta\rlap{-}\iota = 0.0025$ for $[-0.02,0]$, showing the smooth transition of the growth rate and frequency between modes of different dominant helical families.

\begin{figure*}[h!]
\centering
\includegraphics[width=1.0\textwidth]{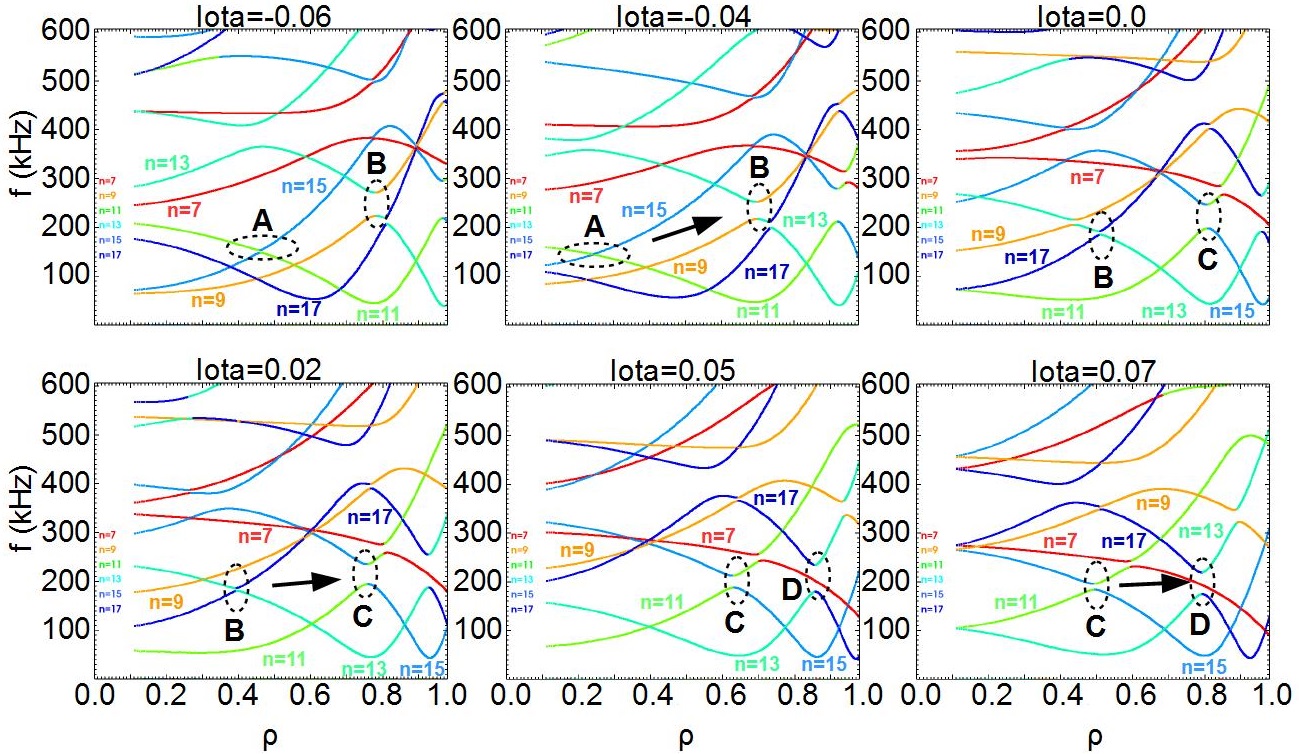}
\caption{Alfv\' en gaps evolution for different $\rlap{-}\iota$ profile displacements: (a) $\Delta\rlap{-}\iota = -0.06$ (b) $\Delta\rlap{-}\iota = -0.04$ (c) $\Delta\rlap{-}\iota = 0.0$ (d) $\Delta\rlap{-}\iota = 0.02$ (e) $\Delta\rlap{-}\iota = 0.05$ (f) $\Delta\rlap{-}\iota = 0.07$. The reference gaps are highlighted by dotted oval lines. The black arrow indicate a transition between different helical families.}\label{FIG:14}
\end{figure*}

Figure~\ref{FIG:14} shows how the Alfv\' en continuum gaps evolve for different $\rlap{-}\iota$ profile displacements, explaining the decrease/increase phases of the AE frequency in Figure~\ref{FIG:12}. From $\Delta\rlap{-}\iota = -0.06$ to $\Delta\rlap{-}\iota = -0.04$ cases, the frequency of the gap A (helical family $n=7,11,15$) decreases, coherent with the frequency drop observed in Figure~\ref{FIG:12}. In case $\Delta\rlap{-}\iota = -0.04$ there is a transition between $n=7,11,15$ to $n=9,13,17$ helical family (see Figure~\ref{FIG:13}), leading to an increase of the AE frequency because the gap B is at a higher frequency than gap A, also observed in Figure~\ref{FIG:12}. From $\Delta\rlap{-}\iota = -0.04$ to $0.0$ case the gap B frequency drops. In $\Delta\rlap{-}\iota = 0.02$ case there is again a transition between helical families, from $n=9,13,17$ to $n=7,11,15$ , leading to an increase of the frequency because gap C frequency is higher. From $\Delta\rlap{-}\iota = 0.02$ to $0.05$ case the gap C frequency drops. Case $\Delta\rlap{-}\iota = 0.07$ shows a transition from $n=7,11,15$ to $n=9,13,17$ helical family, leading to an increment of the AE frequency because the gap D frequency is higher. In summary, the decreasing phase is linked with the evolution of the Alfv\' en gap although the increasing phase is related to the transition between gaps of different helical families.

\begin{figure*}[h!]
\centering
\includegraphics[width=0.8\textwidth]{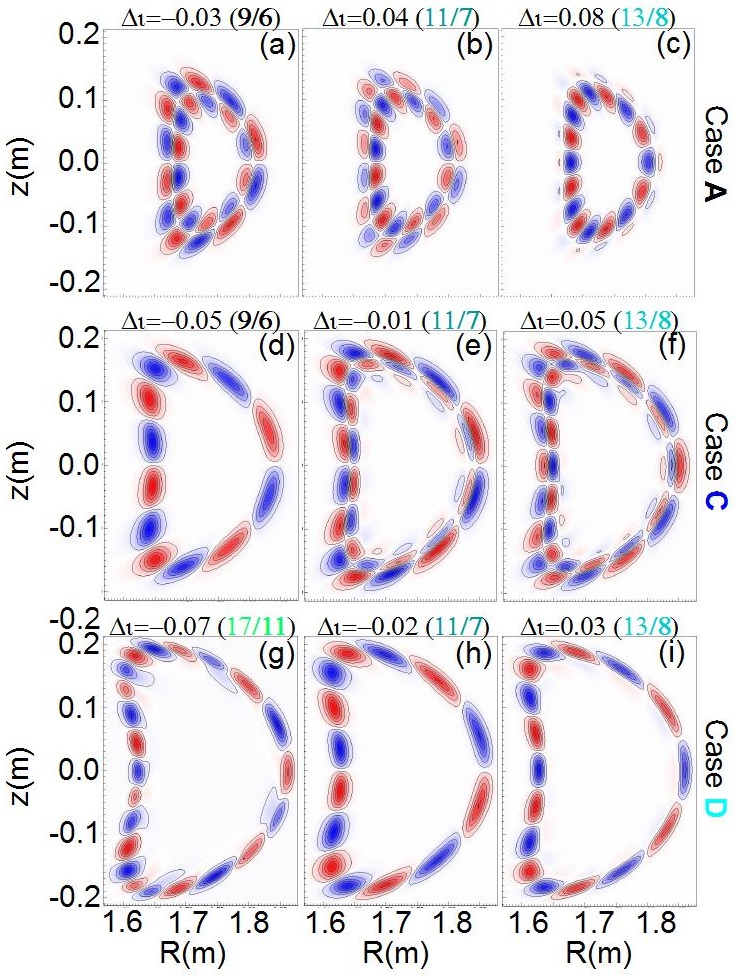}
\caption{2D plots of the $\Phi$ potential for different $\Delta\rlap{-}\iota$ displacements. Case A (panels a to c), case B (panels d to f) and case D (panels g to i). The coordinates are local to the toroidal angle location at the beginning of a field period ($\zeta = 0$).}\label{FIG:15}
\end{figure*}

Figure~\ref{FIG:15} shows the $\Phi$ potential (namely the perturbation) in 2D plots for different $\Delta\rlap{-}\iota$ displacements and dominant modes. The number of the perturbation islands is associated with the dominant mode. In the plasma core, for $\Delta\rlap{-}\iota = -0.03$ (panel a), $0.04$ (b) and $0.08$ (c) the perturbations are driven by the $9/6$, $11/7$ and $13/8$ modes, respectively. A double pattern of islands with opposite parity is observed if the magnetic part is radially even with respect to the rational surface (O-point) and there is a clear dominant mode parity, as in panels (a) and (b), although in panel (c) the double pattern is weaker because the mode $-13/8$ is marginally dominant ($13/-8$ eigenfunction crosses x-axis but the local minima is small compared to the local maxima). If the magnetic part is radially odd the structure is similar to panel (d). The same conclusions can be made for the perturbations in the middle of the plasma for $\Delta\rlap{-}\iota = -0.05$ and the $9/6$ mode in panel (d), $\Delta\rlap{-}\iota = -0.01$ and mode $11/7$ (e) and $\Delta\rlap{-}\iota = 0.05$ and mode $13/8$ (f); also in the plasma periphery for $\Delta\rlap{-}\iota = -0.07$ and mode $17/11$ in panel (g), $\Delta\rlap{-}\iota = -0.02$ and mode $11/7$ (h) and $\Delta\rlap{-}\iota = 0.03$ and mode $13/8$ (i).

\section{Discussion and conclusions} \label{sec:conclusions}

The present study reproduces the most relevant features of TJ-II AE activity driven by the NBI energetic particle component and demonstrates the usefulness of a hybrid reduced MHD/EP Landau closure model for modeling these instabilities. The parametric studies  illustrate the effect of the energetic particle density profile, resonance efficiency and destabilization intensity, as well as the device magnetic field topology, on the AE stability.

The simulations result indicate that including the helical mode couplings in the analysis is mandatory to reproduce TJ-II AE measurements. The simulations with helical mode couplings reproduce AEs in the range of frequencies observed in TJ-II, predicting the destabilization of HAEs in the TJ-II NBI operation regime. The plasma core is unstable to $n=7,11,15$ HAEs with $f = 400$ kHz, destabilized if $\beta_{f} < 0.0025$. In the middle of the plasma, $n=9,13,17$ HAEs with $f=250$ kHz and $n=7,11,15$ EPMs with $f=[130,190]$ kHz are unstable, although the $n=9,13,17$ EPM shows a lower $\beta_{f}$ threshold, smaller than $0.0025$. In the outer plasma, $n=7,11,15$ and $n=9,13,17$ HAEs with $f = 200$ kHz are unstable, showing a similar growth rate and a $\beta_{f}$ threshold, smaller than $0.0025$. For NBI operations with $\beta_{f}$ above the TJ-II NBI operation regime, the $n=7,11,15$ HAEs are still dominant in the plasma core, although in the middle plasma, if $\beta_{f} > 0.01$, $n=7,11,15$ EPMs are dominant. In the plasma periphery, $n=7,11,15$ HAEs with $f = 120$ kHz and $n=9,13,17$ HAEs with $f = 140$ kHz show similar growth rates even if $\beta_{f}$ is above the TJ-II NBI operation regime. 

If the $v_{th,f}/v_{A0}$ ratio is below the TJ-II NBI operational regime, a $11/7$ GAE with $f = 75$ kHz can be destabilized if $v_{th,f}/v_{A0}=0.2$, as well as $11/7$ (middle plasma and periphery) and $17/11$ (plasma core) EPM if $v_{th,f}/v_{A0}=0.1$ with $f < 100$ kHz. If the $v_{th,f}/v_{A0}$ ratio is above the TJ-II NBI operation regime, an $n=7$ TAE with $f=425$ kHz is destabilized.

Table~\ref{Table:3} summarizes the dominant unstable modes as the NBI injection intensity (fixed $v_{th,f}/v_{A0}=0.31$) and voltage changes (fixed $\beta_{f}=0.01$).

\begin{table*}[h]
\centering
\begin{tabular}{c c c c c c}
\hline
$v_{th,f}/v_{A0}$ & $0.1$ & $0.2$ & $0.2-0.3$ & $0.3-0.5$ & $> 0.5$ \\ \hline
 & $11/7$,$17/11$ EPM & $11/7$ GAE & $11/7$,$9/6$,$13/8$ HAEs & $7/4$,$7/5$ TAE & MHD \\ \hline
\end{tabular}
\begin{tabular}{c c c c}
\hline
$\beta_{f}$ & $<0.002$ & $0.002-0.005$ & $>0.005$ \\ \hline
 & $11/7$ EPM, $15/9$ ballooning & $9/6$,$11/7$,$13/8$ HAEs & $11/7$ GAE  \\ \hline
\end{tabular}
\caption{Dominant unstable modes if the NBI injection intensity and voltage is below, above or in the operation regime of TJ-II experiment.} \label{Table:3}
\end{table*}

For the equilibria analyzed, the optimized TJ-II magnetic field topology to enhance AE stability requires a $\rlap{-}\iota$ profile between $[0.845,0.979]$, in order to avoid the destabilization of AEs in the core and middle of the plasma. Full suppression of the AE activity is not achieved in the outer plasma by displacing the $\rlap{-}\iota$ profile, although if $\rlap{-}\iota = [0.945,1.079]$, the growth rate of the AE is 4 times smaller compared to the standard $\rlap{-}\iota$ profile, with a $f = 80$ kHz.

The set of simulations performed for small $\rlap{-}\iota$ profile displacement, $\Delta \rlap{-}\iota = 0.01$, reproduce the sawtooth-like, frequency-sweeping evolution of the AE frequency observed in TJ-II. The frequency oscillations are caused by the destabilization of different helical families modes as the $\rlap{-}\iota$ profile is displaced. In the plasma core, a negative displacement of the $\rlap{-}\iota$ profile leads to a dominant $9/6$ mode, although a positive displacement leads first to a dominant $11/7$ mode followed by a dominant $13/8$ mode for $\Delta \rlap{-}\iota > 0.05$. In the middle of the plasma, a negative displacement of $\Delta \rlap{-}\iota < -0.02$ destabilizes the $17/11$ mode, but if $\Delta \rlap{-}\iota < -0.03$, the $9/6$ mode dominates. A positive displacement of $\Delta \rlap{-}\iota > 0.02$ destabilizes the $13/8$ mode, but if $\Delta \rlap{-}\iota > 0.06$, the mode $15/9$ is dominant. In the outer plasma, a negative displacement $\Delta \rlap{-}\iota < -0.04$ destabilize the $17/11$ mode, but if $\Delta \rlap{-}\iota < -0.06$, the $9/6$ mode is dominant. For a positive displacement $\Delta \rlap{-}\iota > 0.01$, the $13/8$ mode is unstable, but for $\Delta \rlap{-}\iota > 0.04$ the $15/9$ mode is destabilized. The regime of AE frequency increase/decrease can be explained as the evolution of the Alfv\' en gaps (decreasing frequency regime) and the transition between Alfv\' en gaps associated with different helical families (increasing frequency regime).

\ack
This material based on work is supported both by the U.S. Department of Energy, Office of Science, under Contract DE-AC05-00OR22725 with UT-Battelle, LLC. This research was sponsored in part by the Ministerio of Economia y Competitividad of Spain under project no. ENE2015-68265-P. We also want to acknowledge Alexander Melnikov and the TJ-II group at CIEMAT for providing us the initial VMEC equilibria and useful discussions regarding the experimental phenomena. 

\hfill \break

\end{document}